# Interactive Overlays of Journals and the Measurement of Interdisciplinarity on the basis of Aggregated Journal-Journal Citations

Loet Leydesdorff,*[a] Ismael Rafols,[b] & Chaomei Chen [c]

**Abstract**
Using "Analyze Results" at the Web of Science, one can directly generate overlays onto global journal maps of science. The maps are based on the 10,000+ journals contained in the Journal Citation Reports (JCR) of the Science and Social Science Citation Indices (2011). The disciplinary diversity of the retrieval is measured in terms of Rao-Stirling's "quadratic entropy." Since this indicator of interdisciplinarity is normalized between zero and one, the interdisciplinarity can be compared among document sets and across years, cited or citing. The colors used for the overlays are based on Blondel *et al.*'s (2008) community-finding algorithms operating on the relations journals included in JCRs. The results can be exported from VOSViewer with different options such as proportional labels, heat maps, or cluster density maps. The maps can also be web-started and/or animated (e.g., using PowerPoint). The "citing" dimension of the aggregated journal-journal citation matrix was found to provide a more comprehensive description than the matrix based on the cited archive. The relations between local and global maps and their different functions in studying the sciences in terms of journal literatures are further discussed: local and global maps are based on different assumptions and can be expected to serve different purposes for the explanation.

[a] Amsterdam School of Communication Research (ASCoR), University of Amsterdam, Kloveniersburgwal 48, 1012 CX Amsterdam, The Netherlands; loet@leydesdorff.net ; * corresponding author.
[b] SPRU (Science and Technology Policy Research), University of Sussex, Freeman Centre, Falmer Brighton, East Sussex BN1 9QE, United Kingdom; *Ingenio* (CSIC-UPV), Universitat Politècnica de València, València, Spain; i.rafols@sussex.ac.uk.
[c] College of Information Science and Technology, Drexel University, 3141 Chestnut Street, Philadelphia, PA 19104, USA; Chaomei.Chen@cis.drexel.edu .

**Introduction**

Leydesdorff & Rafols (2012) used 2009-data to generate global maps of science at the journal level from the aggregated journal-journal citation data contained in the Journal Citation Reports (JCR) of the *Science* and *Social Science Citation Indices* 2009. In this communication, we follow up on that study by refining the methods and offering maps for 2011 as the most recently available JCRs. Unlike the previous data (for 2009) which were entirely downloaded from the web (with some transmission error), the new maps are based on cleaned data provided by Thomson-Reuters under a license agreement, and are therefore relatively error-free.[1] Whereas the previous study explored different options for the visualization, we use the conclusions of that study and focus here on the visualization using VOSViewer (Van Eck & Waltman, 2010).[2] More recently (July 2012), VOSViewer and Pajek (v3)[3] were further integrated. Furthermore, the new routines no longer require downloading the retrieval, but can operate directly at the interface of Web-of-Science (WoS).

The technique of using overlay maps—which is well-known from its use at Google Maps (e.g., Leydesdorff & Persson, 2010; Bornmann & Leydesdorff, 2011; Leydesdorff & Bornmann, 2012; Leydesdorff, Kushnir, & Rafols, in press)—was introduced into science mapping by Boyack and collaborators in unpublished studies in the mid-2000s (see Boyack, 2009) and elaborated into interactive overlays at the Internet by Leydesdorff & Rafols (2009) and Rafols *et al.* (2010). These latter studies used Web-of-Science (WoS) Subject Categories that are attributed to journals by professional indexers and semi-automatically by computer programs on the basis of criteria such as the content, the title, and the citation patterns of journals (Bensman & Leydesdorff, 2010; Pudovkin & Garfield, 2002: 1113n.). The categories, however, are overlapping and imprecise (Boyack *et al.*, 2005; Rafols & Leydesdorff, 2009; cf. Rafols *et al.*, 2010: 1887).

Visualization of the entire set of 9,162 journals in 2009 was not possible at the time because of computer capacities and the unsolved problem of the cluttering of the many labels in the layout. Local maps of journals, for example, were made in the late 1980s with only the 26 lower-case and 26 upper-case characters of the alphabet as labels using Multidimensional Scaling (MDS) under DOS (e.g., Leydesdorff, 1986).[4] However, the situation was improved when the network analysis software Pajek became available under Windows '96. Pajek allows, among other things, to display labels for specific clusters or partitions. Using the Pajek format, Leydesdorff (2007) brought ego-networks in terms of citations online for mapping all journals contained in the JCRs 2003-2007.[5]

---

[1] The analysis is based on the JCR as made available at the first release by Thomson-Reuters, July 2012. Approximately 90 journals were added in an update in October 2012, and several impact factors were at this occasion revised.
[2] VOSViewer is a program for network visualization freely available at http://www.vosviewer.com .
[3] Pajek is a network visualization and analysis program freely available for non-commercial usage at http://pajek.imfm.si/doku.php?id=download .
[4] The MDS routine MINISSA was integrated into UCINet (v4) in the late eighties.
[5] These maps are available at http://www.leydesdorff.net/jcr04 ; similar maps for the Chinese Science Citation Index CSTPD 2003-2005 are available at http://www.leydesdorff.net/istic04/ (Zhou & Leydesdorff, 2004). Local maps for the Arts & Humanities Citation Index 2008 can be retrieved from http://www.leydesdorff.net/ah08/browse.html (Leydesdorff & Salah, 2010; cf. Leydesdorff, Hammarfelt, & Salah [2011] for global mapping of the A&HCI).

The capacity to print or display labels on a single page or screen is limited. Using more than seventy to one hundred labels, the representation of a network as a map can easily become too crowded as the labels begin to overlap and clutter. Both VOSViewer and Gephi[6] have solved this problem by offering the possibility to foreground certain labels (those with a high value of a given node-attribute) more than others. In Gephi, the label size can be set proportionally to the size of the attribute. The downside of this proportional sizing is that labels of specialist journals can become so tiny that they cannot be read without zooming in (Leydesdorff, Hammarfelt, and Salah, 2011). In VOSViewer, the labels of nodes with small values of the network attribute (e.g., degree centrality) are faded for the sake of readability. However, one can zoom in and then these labels again become readable, or the user can move the cursor to a journal with the mouse, and then bring an otherwise suppressed label to the fore. For our purpose, this functionality is optimal: it solves the problem of visualizing large datasets in cases where the labels contain essential information (in our case, the journal names). The labels are available, but hidden when not needed visually.

Unlike network visualization programs such as Pajek and Gephi, VOSViewer uses an MDS-like algorithm (Kruskall & Wish, 1978) to position the nodes instead of a forced-based spring layout (e.g., Kamada & Kawai, 1989; Fruchterman & Reingold, 1991). The latter algorithms operate to minimize the stress in the sum of individual relations in the graph, whereas MDS (and its derivates) minimizes stress in the *system* of relations under study in terms of the dimensions of the latent structure (Leydesdorff, in press). However, in this study we are interested precisely in the structural dimensions of the journal network at the systems level, and therefore the map of the multi-dimensional vector space (i.e., similarities among citation distributions) will be used instead of the network of individual relations (i.e., citations as valued ties between journals).

The distances separating the journals can be calculated in the vector space on the basis of the cosine similarity—as a proximity measure—between their cited or citing distributions (that is, vectors) of relations, respectively. Unlike the Pearson correlation that normalizes with reference to the mean of the distribution, we use the cosine as a non-parametric proximity measure since citation distributions are highly skewed (Ahlgren *et al*., 2003; Seglen, 1992; cf. Leydesdorff, 1998).

**Methods and data**

The data was harvested from the *Journal Citation Reports* (JCR) 2011 in September 2012. First, the JCRs of the Science and Social Science Editions of this database were merged. On the basis of this data an aggregated journal-journal citation matrix of 10,675 journals was constructed.[7] Of the $10,675^2$ = 113,955,625 cells only 2,207,789 (= 1.94%) are filled with values larger than zero; the grand total of the matrix is 35,295,459 citations, or on average 15.99 per cell with a value larger than zero. The data was gathered from the "citing" side. In the SCI and SSCI, the long tails of low values are sometimes summed up on this (citing) side as "all others". This cutoff at the lower end varies in the JCR with the sizes of the tails. However, since the file contains also 1,226,364 cells with a value smaller than five (55.54% of the non-zero cells), one can expect the remaining inaccuracy because of the data processing to be very small.

---

[6] Gephi is a freeware programs for network analysis and visualization freely available at https://gephi.org/users/download/ .
[7] The Science Edition 2011 contains 8,281 journals, and the Social Science Edition 2011 contains 2,943 journals. Of these journals, 549 are contained in both databases.

The aggregated journal-journal citation matrix was transformed into a cosine-normalized similarity matrix both in the being-cited and the citing directions. Matrices can then be exported in formats that can be read by the various visualization programs. We use SPSS (v.19) for the cosine normalization and Pajek and UCINet for the data manipulation. As noted, VOSViewer is used for the visualizations.[8]

After normalization in terms of the citing patterns, cosine values were larger than zero for 65,349,785 cells (57.34% of $N^2$). With a threshold of cosine > 0.2, the similarity matrix can significantly be reduced to only 3,151,994 (off-diagonal) values larger than zero (2.77%). Of the 10,675 journals, 10,330 (96.8%) are nevertheless connected into the largest component. This largest component is used for the mapping. Visualization software uses largest components because isolated and non-related components cannot be positioned unambiguously with reference to the largest component.

We worked with a standard laptop with 8 GB internal memory under Windows 7, 64-bits. VOSViewer gave no error message for processing the largest component of 10,330 journals. The computation took approximately two hours, but one needs to generate the basemap only once since the coordinates can thereafter be saved and used again. *Mutatis mutandis*, the largest component in the cited direction was 10,256 (96.1%). In this case, three more journals were removed because they generated outlier points, distorting the representation in VOSViewer.

The abbreviation "VOS" in VOSViewer stands for "visualization of similarities." The algorithm used for this is akin to that of MDS: VOSViewer minimizes a stress function at the systems level (Van Eck *et al*., 2010; cf. Kruskall & Wish, 1978; Leydesdorff & Schank, 2008). Waltman *et al*. (2010) have further integrated a clustering algorithm into the program that operates on the basis of the same principles as the positioning of the nodes in the map. The cluster results are automatically colored into the map, but the colors of the clusters can be changed interactively.[9] Additionally, a representation of the map as a density or heat map is provided in VOSViewer.

Eleven clusters were generated in VOSViewer using the citing patterns and the default value for the modularization ($\gamma = 1$; Waltman *et al*., 2010). Leydesdorff and Rafols (2012) used this default solution, but the new version of the maps will be based on modular decomposition using Blondel *et al*.'s (2008) algorithm for the decomposition (in Pajek). This algorithm is more commonly used. Twelve clusters are then distinguished in the citing dimension, and 40 in the cited. Thus, the cited map is finer-grained than the citing one, whereas the citing one is more clearly structured (Figure 1). In a later section of the paper, we will discuss how the user can replace the classification and coloring with any other one—including the one provided by VOSViewer. For reasons of presentation, we also postpone the discussion about the measurement of interdisciplinarity using the overlay maps.

---

[8] Chen & Leydesdorff (in preparation) will make similar functionalities available in CiteSpace.
[9] The clustering algorithm operates with a parameter ($\gamma$) that can be changed interactively in order to generate more or fewer clusters in the solution.

**Construction of the basemaps**

Figure 1 provides the map based on the citing patterns (cosine > 0.2) and using Blondel *et al.*'s (2008) algorithm for the coloring of 12 communities. The resemblance to the maps based on WoS Subject Categories is striking (Leydesdorff, Carley, & Rafols, 2013; Rafols *et al.*, 2010); and this croissant-like structure also accords with Klavans & Boyack's (2009) conclusion that a consensus has increasingly emerged regarding the shape of journal maps based on aggregated citations.

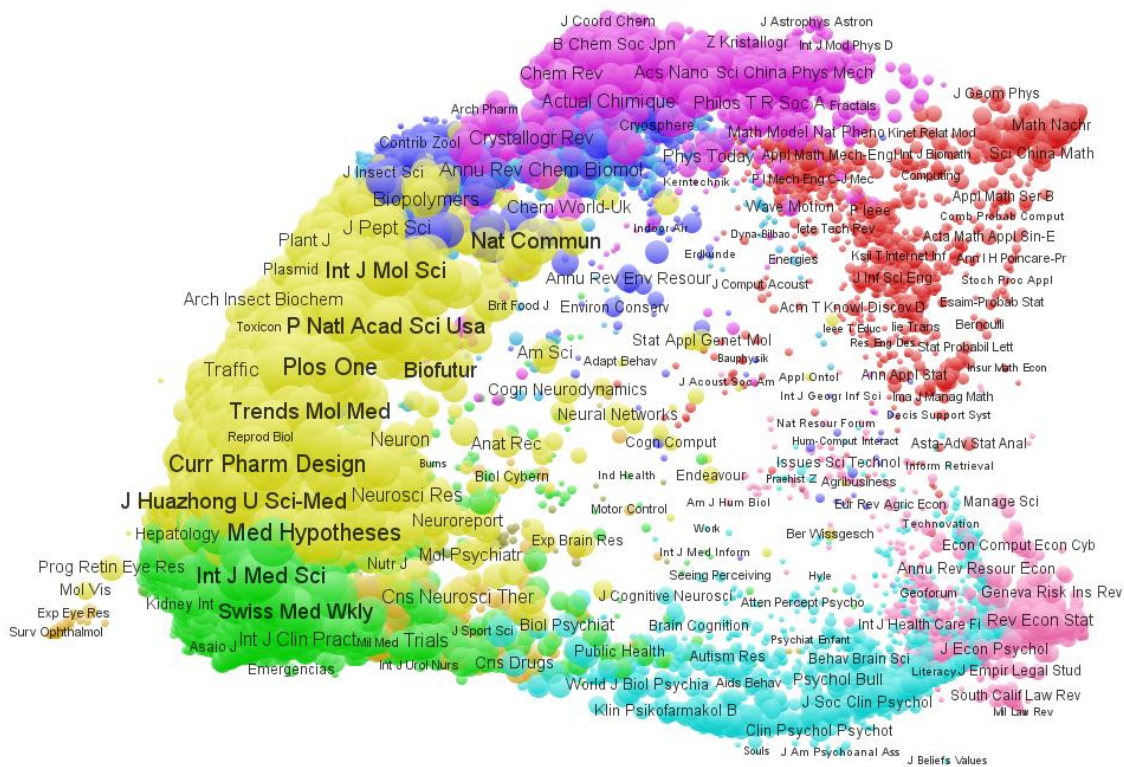

**Figure 1**: 10,330 journals similar in their citing patterns above cosine > 0.2; 12 colors (clusters; $Q = 0.575$).[10] This map can be viewed directly in VOSViewer via WebStart at http://www.vosviewer.com/vosviewer.php?map=http://www.leydesdorff.net/journals11/citing_all.txt&label_size=1.0&label_size_variation=0.3 .

The corresponding figure based on "being-cited" patterns (not shown here) is more compressed because the visibility of relatively isolated groupings in the border regions (that is, peninsulas of the large component) leaves less space for the central grouping. This figure can be web-started at http://www.vosviewer.com/vosviewer.php?map=http://www.leydesdorff.net/journals11/cited_all.txt&zoom_level=1.2&label_size=1.0&label_size_variation=0.3. As noted, 40 clusters are

---

[10] Using single-level refinement (in Pajek): $Q = 0.5747$; multi-level refinement $Q = 0.5750$. The number of clusters is 12 in either case (Blondel *et al.*, 2008).

identified—and therefore differently colored—in this figure ($Q = 0.529$; Blondel *et al.,* 2008).[11] When one enlarges this picture (interactively or by including, for example, the command "zoom_level=1.2", as above), the borders are removed and the resulting picture is not so different from the one based on citing patterns.

This overall agreement between citing and cited patterns[12] does not preclude that citing and cited patterns can be very different when examined in greater detail. Figure 2, for example, shows that core journals in bibliometrics belong "citing" (right-hand panel) very specifically and mono-disciplinarily to a thinly spread set of social-science journals, while they are "cited" (in the left-hand panel) inter-disciplinarily in a more heterogeneous set that the algorithm divides among the social sciences, computer science, and economics in terms of different colors (blue for the social sciences; red for computer science; and green for economics and econometrics).

---

[11] Using single-level refinement (in Pajek): $Q = 0.5297$; multi-level refinement $Q = 0.5294$. The number of clusters in 40 in the latter case (used here) and 41 in the former. In summary, $Q = 0.57$ in the citing and $Q = 0.53$ in the cited dimension: the citing matrix has a slightly higher modularity than the cited, which means that the citing behavior is slightly more organized than what is cited. This conclusion is consistent with the small number of 12 citing clusters compared to the larger number of 40 cited clusters.

[12] Because the two matrices are not of the same size, the computation of a correlation (using, for example,the Quadratic Assignment Procedure (QAP) in UCINet) is not possible.

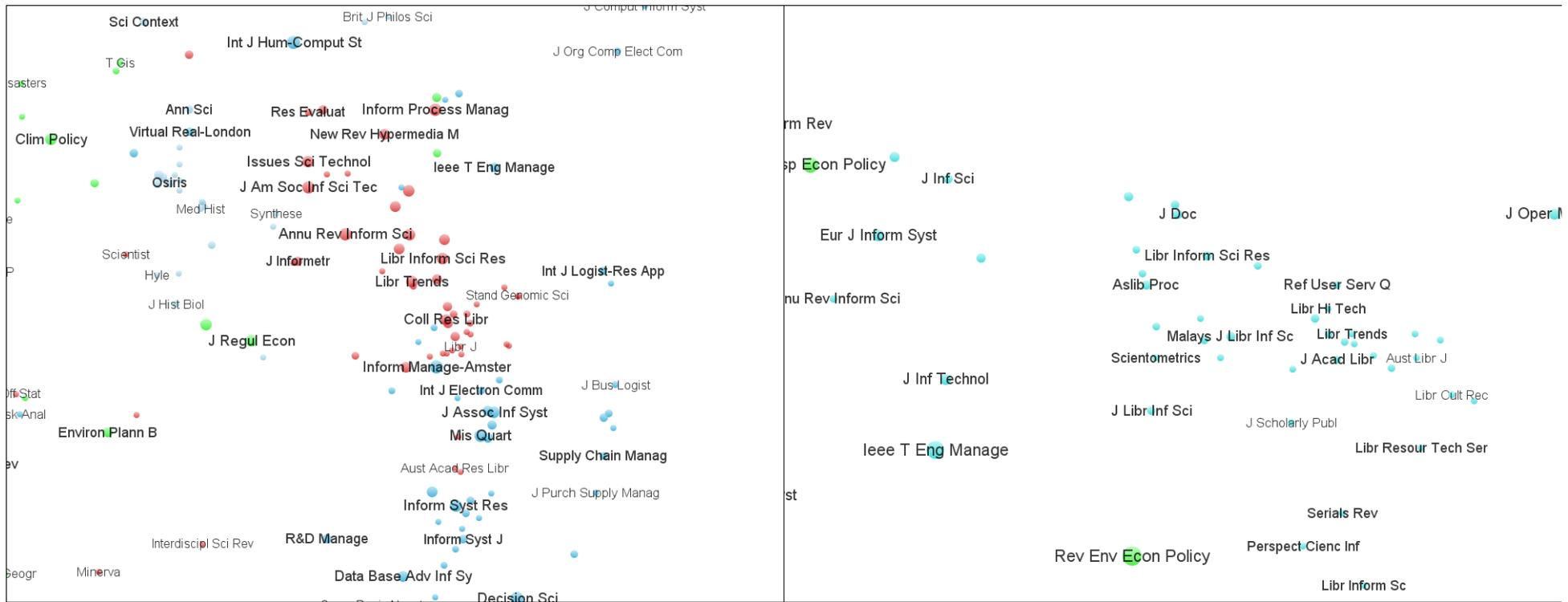

**Figure 2a and b**: Cited (left-side) and citing (right-side) maps of core journals in Library and Information Science; zoomed-in from the general science maps. Economic and econometrics journals are colored green, social-science journals blue, and computer-science journals red.

**The generation of overlay files**

Three programs are made available online for generating overlays. Two of them, "citing.exe" (at http://www.leydesdorff.net/journals11/citing.exe) and "cited.exe" (at http://www.leydesdorff.net/journals11/cited.exe) can process sets of documents downloaded from WoS in the so-called "tagged" format (that is, with labels like "AU " for authors, "TI " for titles, etc.). The third one "analyze.exe" (at http://www.leydesdorff.net/journals11/analyze.exe) uses the results from of the option "Analyze Results" at WoS and does therefore not require downloading the set(s). This option is explained in Appendix A; further instruction can be found at http://www.leydesdorff.net/journals11/index.htm .

In all three routines, the user also needs the table files citing.dbf and/or cited.dbf, respectively, in the same folder; these table files can also be downloaded from the same website at http://www.leydesdorff.net/journals11. In addition to the coordinate information for the maps, the full titles of the journals as provided by JCR are listed in these files. Because there are differences in some cases between the abbreviations in JCR and the *Science Citation Index*, the full titles of the journals are used as keys for the matching. In the case of an unforeseen mismatch—for example, because a journal title was changed—this record will be skipped unless one edits (or duplicates) the title in the corresponding table file.[16]

When the programs and tables are brought into a single folder with the input file—which is downloaded from WoS and renamed "data.txt" in the case of a download or using the default name "analyze.txt" in the case of a listing—an output file can be generated. This file is called either "cited.txt" or "citing.txt" depending on the routine in use. These files can be opened as so-called map-files by VOSViewer. Thereafter, all options commonly available in VOSViewer for the visualization can be used for improving the representations. The resulting figures can be exported as graphic files (.jpg, .png, etc.) or scalable vector graphics (.svg) that can further be edited in InkScape[17] or Adobe Illustrator™. Thus, in addition to the label view (default), one can choose the density view or a heat map; other options make it possible to vary labels in size so that they can be made equally visible, or to change the colors of clusters, etc. In Figure 2 (above), for example, we changed the color for the economics journals from turquoise to green in order to enhance the difference from the blue-colored social-science journals.

As noted, Appendix A provides instruction for the simplest routine (analyze.exe) which does not require a download. The reader is referred to the website for updates and examples. In Appendix B, technical details about the output are provided such as the (default) sizing and coloring of the nodes, the different options for the normalization and clustering, and the possibilities for the user to highlight subsets with different colors or to mark specific journals by intervening in the labeling.[18]

---

[16] Table files in the .dbf format can be read into Excel, but are easier to save after making changes using the spreadsheet editors of OpenOffice or SPSS.
[17] InkScape is freeware available for download at http://inkscape.org.
[18] Chen & Leydesdorff (in preparation) will incorporate multiple views of journal maps in CiteSpace such that different clustering configurations can be viewed and compared.

**Measurement of interdisciplinarity**

The maps enable us to propose an indicator (between zero and one) for the interdisciplinarity of any set downloaded from the Web of Science in terms of the set's distribution across the journals in terms of their distances on the map. These distances can be expressed as a percentage of the maximum distance, that is, the diagonal of the base map. The ratios are then weighted with the proportions of publications in each of the categories (that is, journals) using Rao-Stirling diversity ($\Delta$). This measure is defined as follows:

$$\Delta = \sum_{ij} p_i p_j d_{ij} \tag{1}$$

where $d_{ij}$ is a distance measure between two categories $i$ and $j$, and $p_i$ is the proportion of elements assigned to category $i$—that is, the relative frequency of each journal.

The Rao-Stirling diversity measure was introduced by Rao (1982a and b) and has also been named "quadratic entropy" (Izsák & Papp, 1995) because it measures not only diversity in terms of the spread of the elements among the categories of the classification, but also takes into account the distances among the categories (that is, in this case, among the journals on the map). Stirling (2007, at p. 712) proposed this measure as a general framework for measuring diversity in science, technology, and innovation. Porter *et al*. (2007) also used this measure in their integration score of interdisciplinarity.

Note that this diversity can be considered as a specific—albeit common—operationalization of interdisciplinarity among other possible ones (cf. Barry *et al*., 2008; Klein, 1990; Wagner *et al*., 2011). For example, the concept of "interdisciplinarity" also contains the notion of "intermediation"—which can be operationalized using betweenness centrality (Leydesdorff, 2007; Leydesdorff & Rafols, 2011a)—and "coherence" (Rafols & Meyer, 2010). Using betweenness centrality (an attribute to the nodes of the network), journals can be ranked in terms of their "interdisciplinarity." However, betweenness centrality is defined as a graph-theoretical algorithm and would have to be (re)defined in the vector space on the basis of the cosine-matrix before the network is reduced using cosine > 0.2 as a threshold (Goldstone & Leydesdorff, 2006).

Because betweenness centrality and coherence also vary between zero and one (see Leydesdorff & Rafols [2011b, at p. 856] and Rafols *et al*. [2012, at p. 1286] for the operationalization of coherence), these measures can be added to Equation 1 as additional parameters and thus perhaps generate a more comprehensive measure of "interdisciplinarity." However, Leydesdorff & Rafols (2011a, at p. 96) found different components between betweenness centrality and other (diversity-based) measures of interdisciplinarity using factor analysis of the JCR 2008 as data. In summary, our measure in this study does *not* address the (inter)disciplinary of journals measured in terms of, for example, betweenness centrality, but only the interdisciplinarity of *document sets* measured as Rao-Stirling diversity.

In our opinion, "interdisciplinarity" or its measures (such as diversity) should not be used without specification of the unit of analysis (cf. Wagner *et al*., 2011). In this case, the measure applies only to the interdisciplinarity of downloaded document sets. In a next section, we will extend the options for generating overlays using the journal names in the cited references of these documents, and then specify this as the interdisciplinarity of their respective knowledge bases.[19] Analogously, one can ask for the "interdisciplinarity" of the sets that cite these documents ("the audience set"; Zitt & Small, 2008; cf. Carley & Porter, 2012). A publication set can be monodisciplinary (not diverse in the journals where it is published), but it can be cited interdisciplinarily (by diverse journals) or the other way round. Interdisciplinarity measured as Rao-Stirling diversity can be compared across sets and over time insofar as the same basemaps (cited or citing) are used for the normalization. We shall specify these possible extensions to cited references and citation patterns in a further section (and Appendix C).

In this study, we use the distance on the map $\| x_i - x_j \|$ between each two journals participating in the set as the distance parameter $d_{ij}$ in Eq. 1, as a proportion of the maximally possible distance (that is, the diagonal of the map). This distance measure is an optimization and projection in two dimensions ($x$ and $y$) of the multi-dimensional distances ($1 - cosine$) among journals. Leydesdorff, Kushnir, & Rafols (in press) used the latter measure straightforwardly for an analogous mapping of (USPTO) patents in terms of International Patent Classifications (IPC). However, the number of IPC classes is currently 637, whereas the number of journals is more than 10,000. The number of distances would therefore be on the order of $10^8$. Even after setting the threshold of cosine > 0.2, this number would be on the order of $10^6$, and the size of the files would remain on the order of 50 Mbytes both cited and citing.

Furthermore, the threshold would be too coarse, because more distanced journals may often have a smaller cosine value between them than 0.2, and the variation in the distances ($1 - cosine$) would unnecessarily be reduced from zero to 0.8 given this threshold. Initial explorations led us also to the empirical conclusion that the results would be confounding using the threshold (cosine > 0.2) because of the relative failing of relatedness in interdisciplinary sets above this level of the threshold. By using the distance on the map $\| x_i - x_j \|$ between two journals, these problems are circumvented. Since the MDS-like algorithm of VOSViewer already optimizes in terms of distances, we can use these distances between points directly for the computation of the Rao-Stirling diversity.[20] By normalizing these distances first against the maximum (diagonal) value, one defines the diversity indicator between zero and one (since the $p$-values of the proportions are also fractions of one).

---

[19] This extension assumes that the user has ticked the box within WoS for downloading "cited references" before the downloading, and thus one addresses the interdisciplinarity of another unit of analysis; for example, the knowledge bases of the sets (Bornmann & Marx, 2013; Leydesdorff & Goldstone, 2012).

[20] A related program of VOSViewer, VOSmapping.exe at http://www.vosviewer.com/relatedsoftware/ , allows for specification of the dimensionality to more than two (Ludo Waltman, *personal communication*, December 30, 2012). However, one then loses the relation with the visible distances on the map. Furthermore, the extraction of a third dimension is not expected to add a large percentage to the explanation of the variance in the matrix (Schiffman *et al.*, 1981). Given today's hardware systems limitations, it is not possible to specify the percentages of variance explained by the two first and/or later factors, using SPSS v. 20 (Leydesdorff, 2006). The dimensionality chosen remain therefore a bit arbitrary, unless one were able to use the ($1- cosine$) measure in the $N = 10,330$ dimensions of the full matrix of aggregated journal-journal citations. As noted, this approach would be computationally too intensive given the large value of $N$, but one can pursue such a more precise approach offline. The cosine-normalized Pajek-files, however, are larger than 1 GB before setting a threshold of cosine > 0.2.

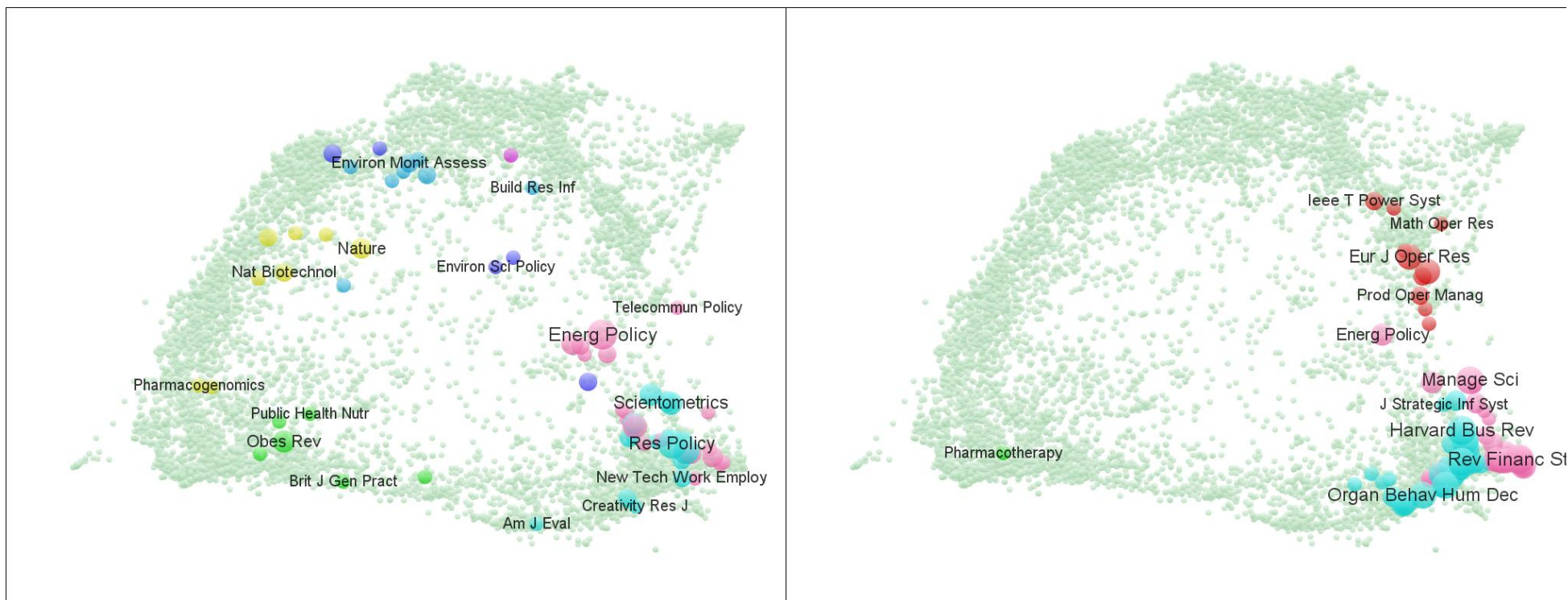

**Figure 3a and b**: Overlay maps 2011 comparing journal publication portfolios from 2006 to 2010 between the Science and Technology Policy Research Unit SPRU at the University of Sussex (on the left; $N$ = 148; available at
http://www.vosviewer.com/vosviewer.php?map=http://www.leydesdorff.net/journals11/fig3a.txt&label_size=1.35)
and the London Business School (on the right; $N$ = 343; available at
http://www.vosviewer.com/vosviewer.php?map=http://www.leydesdorff.net/journals11/fig3b.txt&label_size=1.35).

As an example, we return to the retrieval used by Leydesdorff & Rafols (2012, at p. 328), namely the comparison of the publication portfolios 2006-2010 of the London Business School (LBS) and the Science and Technology Policy Research Unit (SPRU) at the University of Sussex. Using a number of indicators, Rafols *et al*. (2012) showed that the latter unit is far more diverse than the former even though both units are assigned to the same heading of Business & Management in the upcoming UK-wide evaluation, the so-called Research Excellence Framework.

Figure 3 provides the two portfolios of 148 SPRU[21] (to the left) and 343 LBS publications[22] (to the right) as overlays on the 2011 "citing" maps, respectively. Table 1 provides the Rao-Stirling diversities for the two schools in both the cited and citing dimensions.

|  | *Citing* | *Cited* |
|---|---|---|
| *SPRU* ($N = 148$) | 0.218 | 0.136 |
| *LBS* ($N = 343$) | 0.092 | 0.082 |

**Table 1**: Rao-Stirling diversity for 143 SPRU and 343 publications (2006-2010) in both the citing and cited dimensions.

The difference in the values between cited and citing is caused by the larger distances in the citing map when compared with the cited one. The discriminating power of the citing map is therefore larger and the graphs are clearer. Furthermore, "citing" refers to the current knowledge base of the downloaded (sub)set—since citing is the running variable—whereas "cited" refers to the structure in the (cited) archive. We therefore recommend using the routine citing.exe unless one has theoretical reasons for focusing on "cited" or when the more fragmented clustering in the latter map is important for the argument.

The routine provides at each run the value of Rao-Stirling diversity measure on the screen, and this value is saved to a file rao.txt. Note that this file is overwritten in each subsequent run; thus, these values have to be noted separately. Although the coordinates of VOSViewer can vary and take values larger than one (or less than minus one), the diversity values are normalized between zero and one, and the cited *or* citing values in Table 1 may therefore be compared across cases,[23] and one can also compare results using sets of documents for different years. Using PowerPoint, the sequences for different years can also be animated given the stable base map.

**Further extensions to cited and citing sets of documents**

As noted above, one can extend the analysis to the cited and citing sets of documents in terms of other units of analysis—or any unit of analysis that contains full journal names or the conventional abbreviations in the WoS format. For example, all journals titles in documents in WoS (or Scopus, PubMed, etc.) can be matched against the keys contained in the files cited.dbf and citing.dbf that are used for the overlay mapping and the measurement of interdisciplinarity.

---

[21] Of the 155 SPRU papers, 148 were included in the largest component.
[22] Of the 348 LBS papers, 343 were included in the largest component.
[23] Because the map in the "cited" direction is more compressed, however, one cannot directly compare the projections in the cited and citing dimensions in terms of distances.

The two table files contain two keys: the full journal titles and the abbreviated ones using the conventions of WoS for the abbreviations. Our programs automatically correct for variations in upper and lower-case in these titles.

For example, document sets downloaded in WoS contain the abbreviated journal titles in the cited references (field-tag: "CR") in addition to the full journal names of each document which is tagged as "SO" (as an abbreviation of "source"). The journal names in the field "CR" may contain misspellings (Leydesdorff, 2008, Table 4 at p. 285), but Thomson-Reuters has recently invested in v5 of WoS (in 2011) to improve standardization in the CR-field. Two additional (sister) programs are brought online that operate on this field when properly downloaded in the tagged-format and renamed as "data.txt" as above. These routines ("crciting.exe" at http://www.leydesdorff.net/journals11/crciting.exe and "crcited.exe" at http://www.leydesdorff.net/journals11/crcited.exe) operate on the journal names in the cited references in the document set under study using the standard abbreviations of WoS for the comparison (whereas the original programs citing.exe and cited.exe use the full journal titles).[24]

For example, the above used set of 155 documents (Figure 3a) published by authors with an address at SPRU between 2006 and 2010, contains 7,545 cited references of which 2,552 can be matched with the WoS keys for the journal abbreviations. Figure 4 provides the map of the knowledge base of the SPRU authors overlaid on the basemap in the citing direction. The Rao-Stirling diversity is marginally down to 0.214 (from 0.218 in Table 1).[25]

---

[24] The abbreviated journal titles are stored in a file cr.dbf that contains a fieldname "journalcr" used by the routines. When one uses other installations of the *Science Citation Index* such as on Dialog or STN, one may have to rename both the file (to cr.dbf) and this fieldname to "journalcr" (Lutz Bornmann, *personal communication*, 5 January 2013).

[25] CrCited.exe computes a Rao-Stirling diversity of 0.198; that is larger than the value of 0.136 for the document set. It is on our wish-list to develop statistics for the measure such as confidence intervals, but this goal extends beyond the present study.

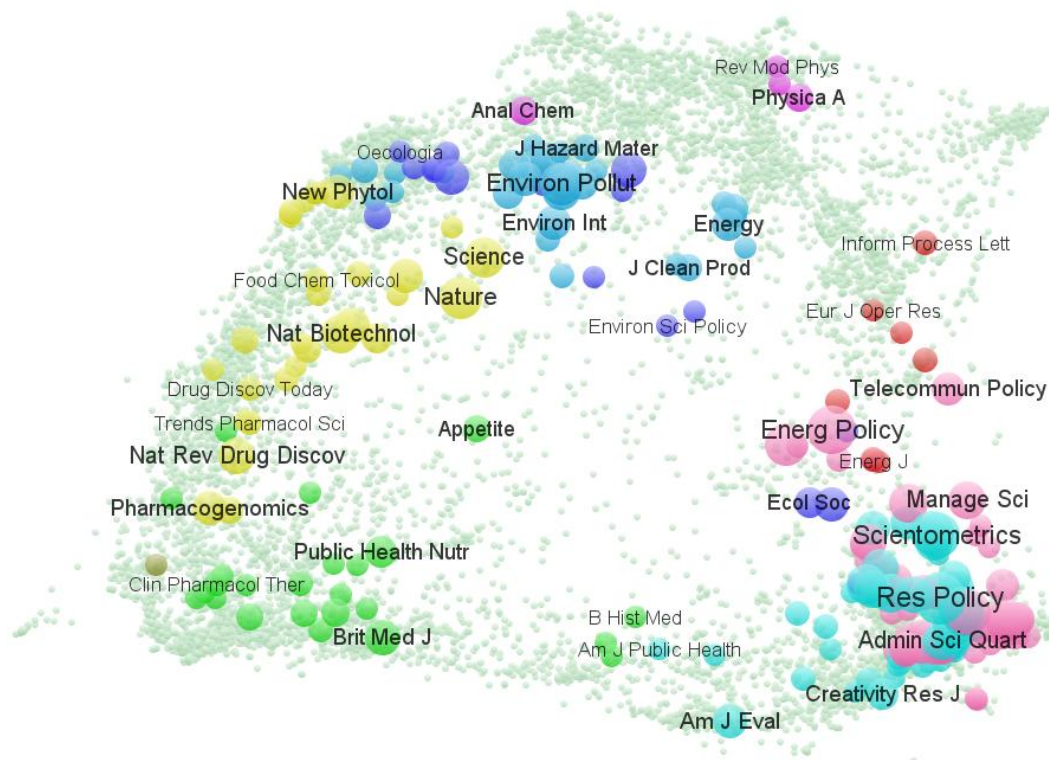

**Figure 4:** 2,552 references in 155 documents published by SPRU authors 2006-2010; mapped as citing patterns; Rao-Stirling diversity is 0.214. This map can be viewed directly in VOSViewer via WebStart at
http://www.vosviewer.com/vosviewer.php?map=http://www.leydesdorff.net/journals11/figure4.txt&label_size =1.35 .

The relatively low rate of matching (2552 of 7545; 33.8%) is perhaps itself indicative of the interdisciplinary nature of these articles, which often result from policy-oriented and externally funded reports. Gibbons *et al.* (1994) called this type of interdisciplinarity the "Mode 2"-type of knowledge production: not only more "interdisciplinarity," but also more engagement with social actors. In the case of the more disciplinarily oriented ("Mode 1") London Business School, 348 documents contain 16,713 cited references, of which 10,034 (60.0%) could be validated in terms of sources at WoS ($\Delta = 0.096$). Thus, the knowledge base of these authors is also more "disciplined" in the sense of being more oriented toward academic objectives.

WoS offers the possibility to download also the citing journals of a set by creating a "Citation Report." The technicalities of using this set of citing journals are provided in Appendix C entitled "Using sets of citing journals from the "Citation Report" of WoS or data from other sources."

**Discussion**

Maps of science in terms of aggregated journal-journal citations provide us with a representation of the intellectual organization of the sciences. Since the maps reflect the aggregated citation choices of authors, the resulting structures are relatively independent from the local actions of authors, editors, publishers, or policy makers. At the level of aggregated journal-journal citations, patterns can be expected to emerge from the aggregation and from interactions among the local actions. These patterns are reproduced and/or change from year to year beyond the control of individual or institutional agency and are in this sense self-organizing at the systems level (Leydesdorff, 1998; Luhmann, 1995). Therefore, one can use journal maps as relatively independent baselines for exploring science dynamics, including the effects of policy interventions and interdisciplinary priority programs (Leydesdorff, 1996; Leydesdorff & Schank, 2008; Leydesdorff, Van den Besselaar, & Cozzens, 1994; cf. Studer & Chubin, 1980, at pp. 269 ff.).

The self-organization of the emerging patterns is caused by layers of selections interacting at lower levels, such as choosing references by authors, refereeing of manuscripts at the journal level, selection among journals in terms of library holdings, and selection of journals by indexing services. The resulting dense areas can be considered as the tips of icebergs in an ocean; the ocean itself is best captured at the article level—for example, using search engines such as Google Scholar. However, the 10,000+ journals with impact factors, and organized in terms of their citation relations into the JCR of the *(Social) Science Citation Index*, represent a high-level control structure which is now fairly well established (Klavans and Boyack, 2009) including even other inputs of field relations such as using click-stream (Bollen *et al*., 2009). Groupings within this structure can be recognized in terms of disciplines and specialties, depending on the zoom of the lens used for the representation.

Global maps of journal networks are in important respects different from local maps. The nesting is not in terms of individual relations in the network, but in terms of functionalities in the vector space (Simon, 1962; Leydesdorff, in press): scholarly journals can belong to (one or more) specialties, and specialties are organized into (sub)disciplines, etc. The decomposition in a function space is different from decomposition in the design space (Bradshaw & Lienert, 1991). The design space is shaped in terms of relations and can be represented as a (co-)occurrence matrix or graph, but the functions operate in a multi-dimensional vector space that can be captured using cosine values among vectors as proxies. Decomposition of this topology is different from the decomposition in the case of only aggregated relations.

Although the vectors of the grand matrix contain mostly zeros (that do not contribute to the cosine), the matrix is in important respects not decomposable: both within journals (Boyack & Klavans, 2011) and at the level of specialties (Leydesdorff & Goldstone, 2012; Rafols & Leydesdorff, 2009) "interdisciplinarity" can be expected to counteract the decomposability of the structures. Sub-matrices represented in local maps, however, can reveal latent structures at a more informed and finer-grained level than global maps. In the case of journal maps, the two options of global and/or local mapping are available as different representations (cf. Klavans & Boyack, 2011).

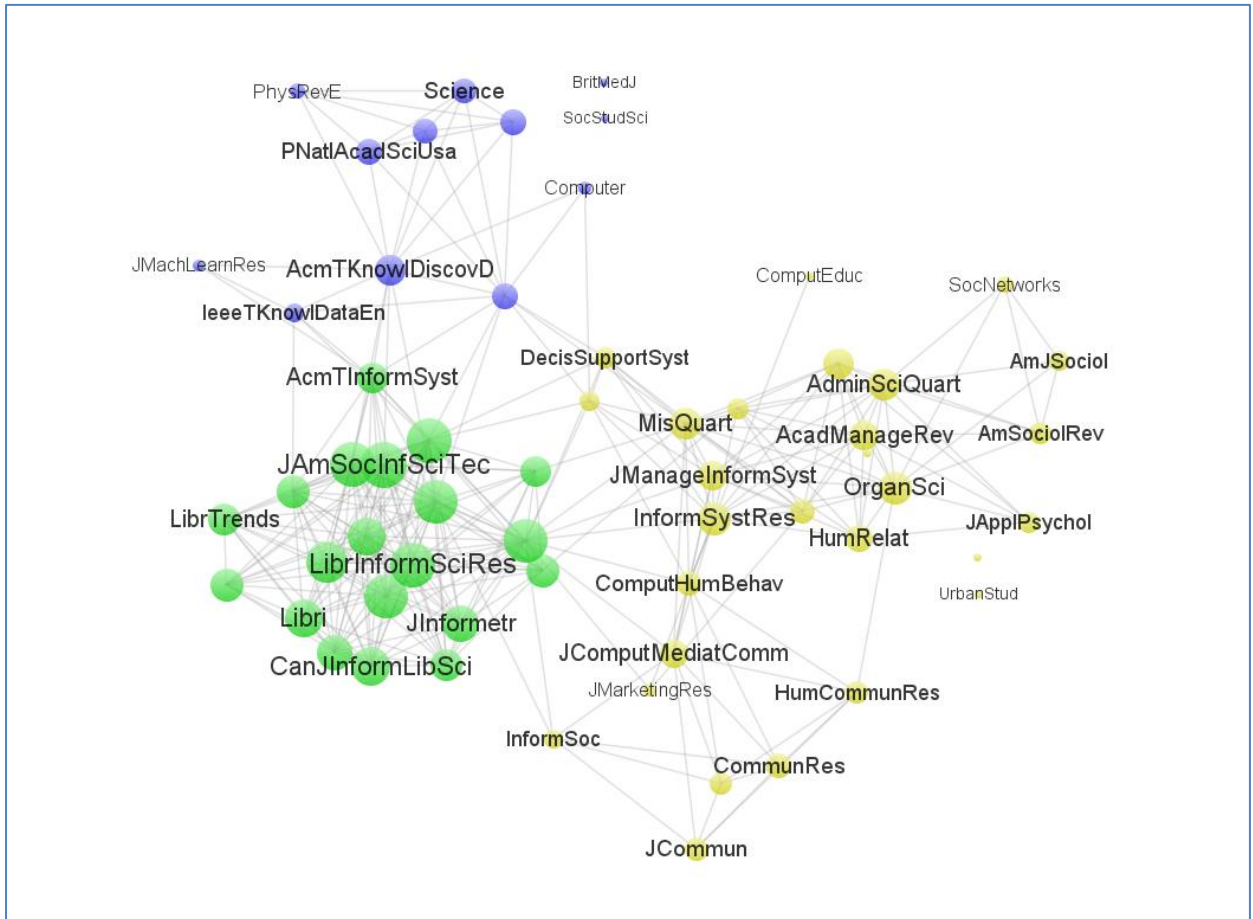

**Figure 5**: Local map of 60 journals most frequently cited by the *Journal of the American Society for Information Science and Technology* in 2011; cosine > 0.2; *Q* = 0.389; 3 communities. This map can be viewed directly in VOSViewer via WebStart at
http://www.vosviewer.com/vosviewer.php?map=http://www.leydesdorff.net/journals11/fig5map.txt&network=http://www.leydesdorff.net/journals11/fig5net.txt&n_lines=3000&label_size=1.35.

Figure 5, for example, shows a local map that can be compared with the zoom of the global map in the right-hand panel of Figure 2 above. In this case, the local environment of *JASIST* is mapped in the citing dimension, otherwise strictly analogous to the global map. Sixty journals were cited by articles in *JASIST* ten or more times in 2011, and are mapped here in terms of their mutual citing patterns. As before, Blondel *et al.*'s (2008) algorithm was used for the grouping and VOSViewer for the visualization after exportation of the network and partitions from Pajek. The figure shows the relevant environments in terms of a core set of journals in library & information science (LIS; green colored), information-systems journals with related environments in yellow, and computer-science and general-science journals in blue. These three groups can be further refined using factor analysis (e.g., Leydesdorff & Cozzens, 1993; Van den Besselaar & Leydesdorff, 1996).

In contrast, the right-hand panel of Figure 2 (above) shows a thin network at the global level because most of these journals are specialized journals that play minor roles in the global network of science that includes large journals in the natural and life sciences. In the context of a global map, the LIS group was attributed to the social sciences, the information-systems group to economics, and the other (sometime multi-disciplinary) journals color the palette of this global environment. In our opinion, the global map is not an appropriate tool for studying specialty structures as latent constructs in maps of science because the visible distinctions tend to remain too coarse. Local maps enable us to study specialty structures in considerable detail.

The global map provides the overview and enables us by using the overlay technique to assess the interdisciplinarity among potentially disparate sets quantitatively. Note that these document sets can be retrieved from WoS on the basis of very heterogeneous search strings including authorships, title words, institutional addresses, publication years, etc., and with or without Boolean operators; for example, for evaluation purposes. The proposed technique does not indicate the interdisciplinarity of journals or specialties (cf. Leydesdorff & Goldstone, 2012), but of sets of documents given the organization of journals on a global map.

**Conclusion**

The journal map based on a cosine-normalized matrix and using the MDS-like solution of VOSViewer captures journals as positions in a vector space that is reduced to the two dimensions of the plane. The first two main dimensions of the underlying citation matrix can be expected to capture the major part of the variance in the matrix (Schiffman *et al*., 1981; see footnote 22 above). However, a map remains a projection (in two dimensions). Unlike spring-embedded solutions, the projection of MDS is not dependent on a seed, but the *system* of journal-citations is projected deterministically.[26] The journals are positioned in the vector space on the basis of the aggregates of their mutual relations (Leydesdorff, in press).

The journal is a more precise unit of analysis when compared with the journal grouping using WOS Categories (Rafols *et al*., 2010; Leydesdorff & Rafols, 2012; Leydesdorff, Carley, & Rafols, in press). The WOS Categories are both divisive and overlapping, since journals can be attributed to several categories, on the one hand, but the cuts between categories remain sharp, on the other. The consequent error reflects uncertainty in the networks about the delineations (Rafols and Leydesdorff, 2009; Rafols *et al*., 2010). In a lower-level networked system of journal, such decisions are not needed since all cosine-normalized distances among journals can be introduced concurrently into the computation. (The threshold of cosine > 0.2 was set above because of technical limitations.)

A network system at the article level would be even more precise, but dysfunctional in terms of overlay files for studying sets, for example in terms of their interdisciplinarity, because the journals are no longer considered as relevant categories. One would be able to position articles, but one cannot position other articles in terms of a baseline of articles. Another advantage of positioning papers in terms of locations on the map of journals is the availability of network measures of interdisciplinarity. Rao-Stirling diversity measure of interdisciplinarity, for example, operates directly on the values that are visible on the map, that is, the distances between the nodes and the (logarithmically normalized) sizes of the nodes given the document set(s) under study.

---

[26] VOSViewer uses a seed and may therefore remain in a local optimum (Van Eck & Waltman, 2012, p. 2; Waltman, *personal communication,* January 8, 2013).

**Acknowledgement**
We would like to thank Nees Jan van Eck and Ludo Waltman for suggestions and comments, and are grateful to Thomson-Reuters for access to the data. We acknowledge support by the ESRC project 'Mapping the Dynamics of Emergent Technologies' (RES-360-25-0076).

**Appendix A**

**Generating journal maps of science at WoS without downloading the sets**

After entering one's search results at WoS, one can click at the top of the screen to the right on "Analyze Results". In the "Results Analysis" that then opens as a next screen, one selects "Source Titles". Thereafter select "All data rows"—different from the default option—and save the results to the file analyze.txt (default). Analyze.exe (at http://www.leydesdorff.net/journals11/analyze.exe) reads this file as default, and generates output files if the file "citing.dbf" and/or "cited.dbf" is also present. "Cited.dbf" is automatically used if the file "citing.dbf" is absent.

Note that analyze.txt is not overwritten in case of a next run, but a new file analyze(1).txt (etc.) is generated by WoS. Rename this file locally into analyze.txt before further processing it. Analyze.exe overwrites files from previous runs.

**Appendix B**

**Coloring, sizing, clustering, and labeling the maps; options for the user**

Our routines set the sizes of the nodes equal to the $\log_4(n + 1)$. The value of *n* is augmented by one in order to prevent the disappearance of a node in the case of a single publication (since $\log(1) = 0$). The base 4 for the logarithm was chosen for pragmatic and esthetic reasons. Depending on the relative sizes, the user may wish to use a function other than the logarithm. The values of *n*, for example, can be retrieved from the table file named "overlay.dbf" which is generated at each run; this file can be read into Excel. By replacing the column labeled "normalized weight" in the map-file (cited.txt or citing.txt) with the values in the column *NPubl* in the file overlay.dbf, for example, one can obtain a map which exhibits a linear relation between the sizes of nodes and their respective publication volumes.[27]

In VOSviewer, one can choose between weighted sizes (normalized by dividing all weights by the average weight) or "normalized weight," that is, using the weights as already normalized by the user. Default output of our routines contains the label "normalized weight"; that is, the base-4 logarithm of the number of papers is used for the sizing of the nodes. This constant normalization enables the user to compare across overlays and to animate them for different years. By removing the word "normalized" from the header of the map-files of VOSViewer (i.e., by replacing the header with "weight"), however, the resulting figures can be esthetically optimized for each dataset independently using the normalization of VOSViewer. The user can change these column headings in the first lines of the files "citing.txt" and "cited.txt" after running the programs citing.exe or cited.exe, but before importing these (map) files into VOSViewer.

If one wishes to assume another classification as the *default* for generating overlays, one has to change the cluster indication in the column named "Blondel" in the tables cited.dbf and/or citing.dbf in this respect (in Excel or SPSS) and save these files thereafter again as .dbf tables with the same name.[28] These table files contain the clustering results of VOSViewer (for the default value of $\gamma = 1$; cf. Waltman *et al*., 2010) in the column headed "cluster" that can be copied to the column with the header "Blondel". Our programs use standardly the values provided in this latter column. The files citing.txt and cited.txt can also be edited, and then the last columns with cluster numbers that dictate the coloring and/or the labels can also be changed specifically using a text editor (or Excel); for example, if one wishes to highlight a specific group by using a different color or, for example, add a marker to specific labels. As noted, one can also export the maps as .svg and then edit the labels using, for example, InkScape[29] or Adobe Illustrator™.

---

[27] The files cited.txt or citing.txt are "comma-separated variable" files (.csv) that can be read and saved, for example, by Excel.
[28] Exporting files in the .dbf format may not be easy in newer versions of Excel, but it is possible using the same spreadsheet in Open Office (or using other programs, including SPSS).
[29] InkScape is freeware available for download at http://inkscape.org.

**Appendix C**

**Using sets of citing journals from the "Citation Report" of WoS or data from other sources**

WoS offers the possibility to download also the citing journals of a set by creating a so-called "citation report." This screen, however, allows only for downloading as comma-separated variables or Excel sheets. The download contains the names of the journals, but not the cited references. After downloading, one can change the field-name in the Excel sheet into "SO" (as in the tagged format) and saving the file using the name "core.dbf", citing.exe and cited.exe will use this file as source information *in the absence of data.txt*, and thus produce the overlay files citing.txt or cited.txt, respectively, and Rao-Stirling diversity values. (As noted, saving an Excel sheet in the .dbf format is easier in OpenOffice or SPSS than in more recent versions of Excel.)

In general, when full journal names are used in external files, the file containing these names have to be named "core.dbf" and the field containing the titles "so;" if one uses the cited references within documents the file has to be named "cr.dbf" and the field with journal name abbreviations "journalcr." In these cases of external files, one answers "no" on the first question of the routines (citing.exe or crciting.exe) asking for the file "data.txt" and the programs will use these externally imported files instead for generating the map-files and the computation of Rao-Stirling diversity, but only insofar as the titles or abbreviations precisely match. One may sometimes have to edit the input in order to obtain a 100% matching; for example, when data sets from Scopus or Google Scholar are used.